\documentclass[twocolumn, twocolappendix]{aastex631}
\usepackage{xspace}	
\usepackage{amsmath}

\newcommand{\arcsecs}{{$^{\prime\prime}$}\xspace}

\newcommand{\sw}[1]{\texttt{#1}}

\received{}
\revised{}
\accepted{}
\submitjournal{ApJ Letters}

\shorttitle{Hosts of Galaxy Cluster FRBs}
\shortauthors{Sharma et al.}

\begin{document}

\title{Deep Synoptic Array science: A massive elliptical host among two galaxy-cluster fast radio bursts}

\correspondingauthor{Kritti Sharma}
\email{kritti@caltech.edu}

\author[0000-0002-4477-3625]{Kritti Sharma}
\affiliation{Cahill Center for Astronomy and Astrophysics, MC 249-17 California Institute of Technology, Pasadena CA 91125, USA.}

\author{Jean Somalwar}
\affiliation{Cahill Center for Astronomy and Astrophysics, MC 249-17 California Institute of Technology, Pasadena CA 91125, USA.}

\author{Casey Law}
\affiliation{Cahill Center for Astronomy and Astrophysics, MC 249-17 California Institute of Technology, Pasadena CA 91125, USA.}
\affiliation{Owens Valley Radio Observatory, California Institute of Technology, Big Pine CA 93513, USA.}

\author{Vikram Ravi}
\affiliation{Cahill Center for Astronomy and Astrophysics, MC 249-17 California Institute of Technology, Pasadena CA 91125, USA.}
\affiliation{Owens Valley Radio Observatory, California Institute of Technology, Big Pine CA 93513, USA.}

\author{Morgan Catha}
\affiliation{Owens Valley Radio Observatory, California Institute of Technology, Big Pine CA 93513, USA.}

\author{Ge Chen}
\affiliation{Cahill Center for Astronomy and Astrophysics, MC 249-17 California Institute of Technology, Pasadena CA 91125, USA.}

\author{Liam Connor}
\affiliation{Cahill Center for Astronomy and Astrophysics, MC 249-17 California Institute of Technology, Pasadena CA 91125, USA.}

\author{Jakob T. Faber}
\affiliation{Cahill Center for Astronomy and Astrophysics, MC 249-17 California Institute of Technology, Pasadena CA 91125, USA.}

\author{Gregg Hallinan}
\affiliation{Cahill Center for Astronomy and Astrophysics, MC 249-17 California Institute of Technology, Pasadena CA 91125, USA.}
\affiliation{Owens Valley Radio Observatory, California Institute of Technology, Big Pine CA 93513, USA.}

\author{Charlie Harnach}
\affiliation{Owens Valley Radio Observatory, California Institute of Technology, Big Pine CA 93513, USA.}

\author{Greg Hellbourg}
\affiliation{Cahill Center for Astronomy and Astrophysics, MC 249-17 California Institute of Technology, Pasadena CA 91125, USA.}
\affiliation{Owens Valley Radio Observatory, California Institute of Technology, Big Pine CA 93513, USA.}

\author{Rick Hobbs}
\affiliation{Owens Valley Radio Observatory, California Institute of Technology, Big Pine CA 93513, USA.}

\author{David Hodge}
\affiliation{Cahill Center for Astronomy and Astrophysics, MC 249-17 California Institute of Technology, Pasadena CA 91125, USA.}

\author{Mark Hodges}
\affiliation{Owens Valley Radio Observatory, California Institute of Technology, Big Pine CA 93513, USA.}

\author{James W. Lamb}
\affiliation{Owens Valley Radio Observatory, California Institute of Technology, Big Pine CA 93513, USA.}

\author{Paul Rasmussen}
\affiliation{Owens Valley Radio Observatory, California Institute of Technology, Big Pine CA 93513, USA.}

\author{Myles B. Sherman}
\affiliation{Cahill Center for Astronomy and Astrophysics, MC 249-17 California Institute of Technology, Pasadena CA 91125, USA.}

\author{Jun Shi}
\affiliation{Cahill Center for Astronomy and Astrophysics, MC 249-17 California Institute of Technology, Pasadena CA 91125, USA.}

\author{Dana Simard}
\affiliation{Cahill Center for Astronomy and Astrophysics, MC 249-17 California Institute of Technology, Pasadena CA 91125, USA.}

\author{Reynier Squillace}
\affiliation{Cahill Center for Astronomy and Astrophysics, MC 249-17 California Institute of Technology, Pasadena CA 91125, USA.}

\author{Sander Weinreb}
\affiliation{Cahill Center for Astronomy and Astrophysics, MC 249-17 California Institute of Technology, Pasadena CA 91125, USA.}

\author{David P. Woody}
\affiliation{Owens Valley Radio Observatory, California Institute of Technology, Big Pine CA 93513, USA.}

\author{Nitika Yadlapalli}
\affiliation{Cahill Center for Astronomy and Astrophysics, MC 249-17 California Institute of Technology, Pasadena CA 91125, USA.}

\collaboration{200}{(The Deep Synoptic Array team)}

\begin{abstract}

The stellar population environments associated with fast radio burst (FRB) sources provide important insights for developing their progenitor theories. We expand the diversity of known FRB host environments by reporting two FRBs in massive galaxy clusters discovered by the Deep Synoptic Array (DSA-110) during its commissioning observations. FRB 20220914A has been localized to a star-forming, late-type galaxy at a redshift of 0.1139 with multiple starbursts at lookback times less than $\sim$3.5~Gyr in the Abell 2310 galaxy cluster. Although the host galaxy of FRB 20220914A is similar to typical FRB hosts, the FRB 20220509G host stands out as a quiescent, early-type galaxy at a redshift of 0.0894 in the Abell 2311 galaxy cluster. The discovery of FRBs in both late and early-type galaxies adds to the body of evidence that the FRB sources have multiple formation channels. Therefore, even though FRB hosts are typically star-forming, there must exist formation channels consistent with old stellar population in galaxies. The varied star formation histories of the two FRB hosts we report indicate a wide delay-time distribution of FRB progenitors. Future work in constraining the FRB delay-time distribution, using methods we develop herein, will prove crucial in determining the evolutionary histories of FRB sources.

\end{abstract}


\section{Introduction} \label{sec:introduction}

Characterizing the stellar population in the neighborhood of extragalactic transients can unveil the nature of their progenitors. The morphology, color, metallicity, age, and star formation history of the host galaxies of supernovae helped constrain their numerous explosion channels~\citep{2020MNRAS.499.1424H, 2014MNRAS.438.1391P, 2010MNRAS.405...57S, 2022ApJ...927...10I}. The hunt for correlations with the host galaxy's stellar mass and metallicity~\citep{2014ApJ...789...23K}, studies of nucleus-offset distribution~\citep{2002AJ....123.1111B}, and ongoing recent star formation~\citep{2016ApJ...817..144B} revealed that the progenitors of long gamma-ray bursts have a short lifetime, prefer dense and low-metallicity stellar environments, and are likely to be found in young starbursts of blue star-forming galaxies with high specific star formation rates~\citep{2016SSRv..202..111P, 2010AJ....139..694L}. Similar studies for short gamma-ray bursts revealed that the hosts are more luminous and found in less actively star-forming regions than long gamma-ray bursts~\citep{2009ApJ...690..231B}. The large nucleus-offsets suggested that short gamma-ray burst progenitors migrate from stellar nurseries to explosion sites, thus hinting towards kicks during the merger of compact object binaries~\citep{2013ApJ...776...18F, 2022ApJ...940...56F}.

The studies of fast radio burst (FRB) host galaxies, enabled by arcsecond-scale localization by modern radio interferometers, have attempted to solve the long-standing mystery of these energetic, short-duration enigmatic explosions~\citep{2023arXiv230205465G, 2022AJ....163...69B, 2021ApJ...917...75M, 2020ApJ...903..152H}. The major conclusions from such studies have been actively incorporated into proposed progenitor models~\citep{2022A&ARv..30....2P, 2019A&ARv..27....4P}. For example, the association of FRB 20121102 with a dwarf, rapidly star-forming galaxy, and a persistent radio source suggested a young magnetar progenitor~\citep{2017ApJ...834L...7T, 2017Natur.541...58C, 2015arXiv151109137K}. However, the discovery of a repeating FRB 20200120E associated with a globular cluster of M81 indicated that the progenitor was formed in a compact binary coalescence event~\citep{2022Natur.602..585K, 2021ApJ...910L..18B}. The diagnostics such as inferred local environments, galaxy types, and accurately derived physical properties of a large sample of host associations can help disentangle the proposed progenitor theories and differentiate FRBs from other extragalactic transients~\citep{2022A&ARv..30....2P}. These studies can determine if FRBs may be formed via one or multiple progenitor channels since FRBs have been found in a spectrum of environments, including dwarf galaxies~\citep{2017ApJ...843L...8B, 2022arXiv221116790B}, spiral galaxies~\citep{2021ApJ...917...75M, 2021ApJ...919L..23F, 2021ApJ...908L..12T, 2020Natur.577..190M}, and globular cluster~\citep{2022Natur.602..585K, 2021ApJ...910L..18B}. The existing sample of host galaxies of FRBs suggests that they are generally star-forming~\citep{2023arXiv230205465G}. The distribution of stellar properties of FRB hosts has been found to be inconsistent with that of long gamma-ray bursts and superluminous supernovae, with a probable analogy with magnetars formed in core-collapse supernovae~\citep{2021ApJ...907L..31B, 2021A&A...656L..15P}. 

Motivated by such studies, in this article we present a detailed analysis of two new FRBs, FRB 20220914A and FRB 20220509G, both of which are located within massive galaxy clusters~(Connor et al., in prep.). While the host galaxy of FRB 20220914A is a star-forming galaxy with a bursty star formation history, the host galaxy of FRB 20220509G is the first early-type quiescent FRB host. In \S~\ref{sec:observations}, we discuss Deep Synoptic Array (DSA-110)\footnote{\url{https://deepsynoptic.org}} detection of these two FRBs and the optical data obtained for their host galaxies. We present our analysis framework and derived galaxy properties in \S~\ref{sec:analysis_framework}. We then compare our FRBs with the existing sample of localized FRBs, the galaxy population, and other extragalactic transients, along with the first attempt to formulate, model, and constrain their delay-time distribution in \S~\ref{sec:frb_progenitors}. We discuss the implications of our results and summarize the article in \S~\ref{sec:discussion}. Throughout, we adopt the Planck13 cosmology~\citep{2014A&A...571A..16P}, where Hubble constant H$_0 = 67.8~\mathrm{km}~\mathrm{s}^{-1}$ Mpc$^{-1}$, cosmological constant $\Omega_{\Lambda} = 0.69$ and matter-density parameter $\Omega_{\mathrm{m}} = 0.31$.

\section{Observations} \label{sec:observations}

\begin{figure*}
    \includegraphics[width=\textwidth]{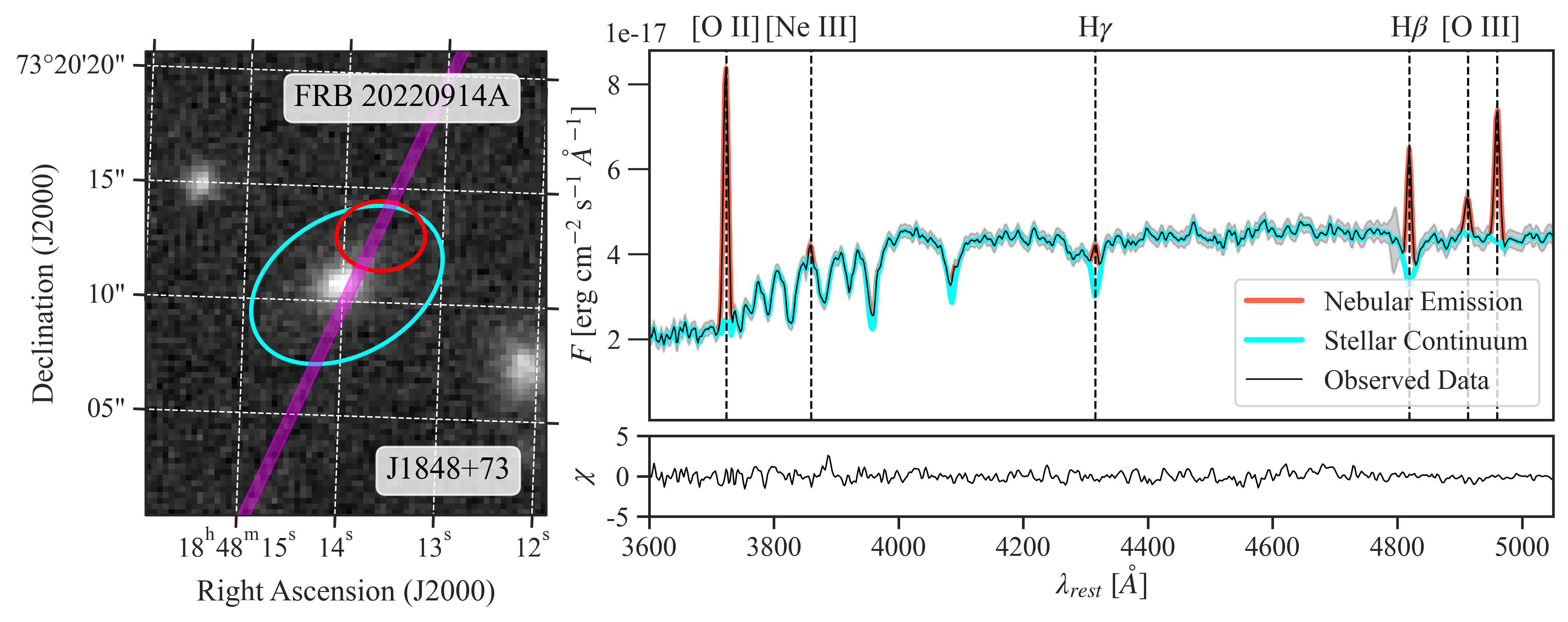}
    \includegraphics[width=\textwidth]{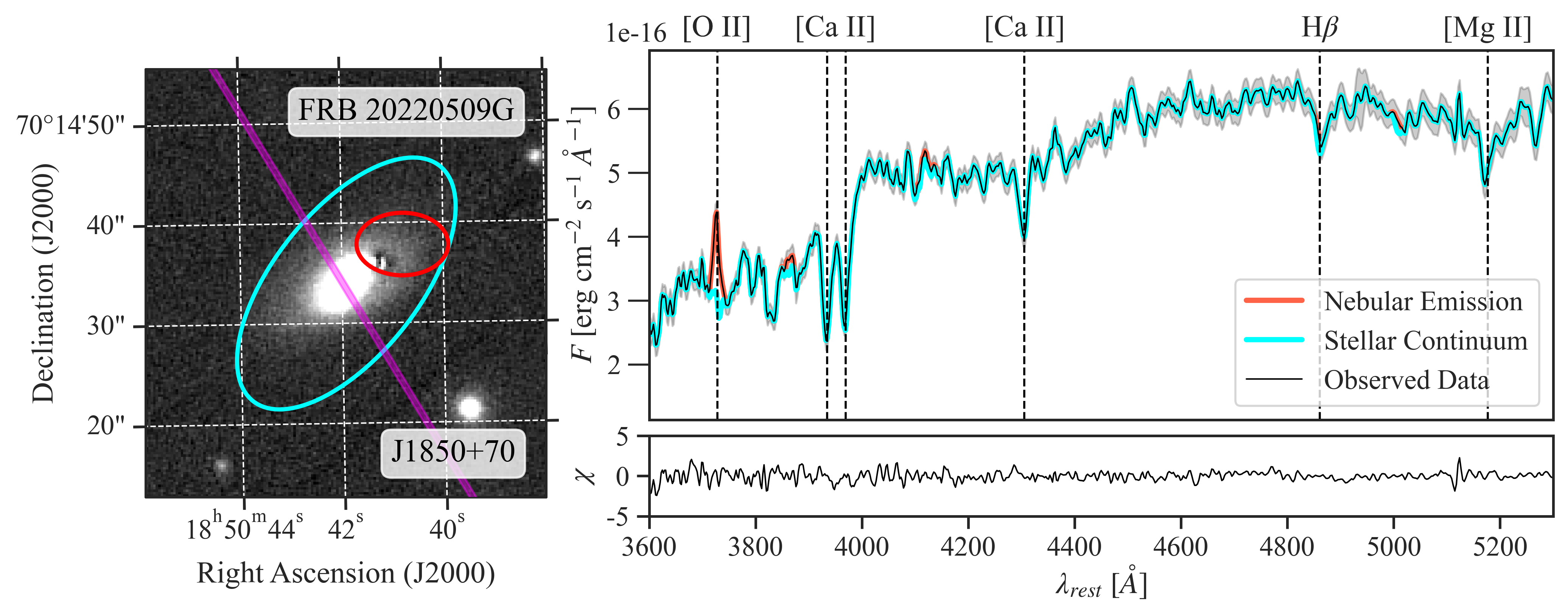}
    \caption{Optical data for the host galaxies of FRB 20220914A (top row) and FRB 20220509G (bottom row). The flux-conservative isophote for photometry (cyan), the slit positions for spectroscopy with Keck-I/LRIS (magenta) and 90\% confidence localization region (red) of both the FRBs overplotted on the PS-1 i-band images are displayed in the left panels. The \sw{pPXF} fits to the stellar continuum (cyan) and nebular emission (red) with corresponding residuals in our Keck-I/LRIS optical spectra (black) of both the host galaxies are included in the right panels.}
    \label{fig:hosts_ppxf_fits}
\end{figure*}

We focus this section on the optical follow-up observations of the host galaxies of FRB 20220509G and FRB 20220914A. A description of the DSA-110 discovery and radio properties of these FRBs is presented in a companion article~(Connor et al., in prep.). FRB 20220509G was localized to (R.A. J2000, decl. J2000) = 18h50m40.8s, $+$70d14m37.8, with a 90$\%$ error ellipse with axes 4.7\arcsec~and 3.2\arcsec~in R.A. and declination respectively. FRB 20220914A was localized to (R.A. J2000, decl. J2000) = 18h48m13.63s, $+$73d20m12.89s, with a 90$\%$ error ellipse with axes 2.0\arcsec~and 1.6\arcsec~in R.A. and declination respectively. The localization procedures were identical to those described in \citet{2023arXiv230101000R}. With regards to the radio properties, it is particularly noteworthy that while no polarized signal or scattering was detected from FRB 20220914A, FRB 20220509G shows evidence for temporal broadening due to scattering with a timescale of 80$\pm$20\,$\mu$s at 1498.75\,MHz, and a Faraday rotation measure of $-111.54\pm1.50$\,rad\,m$^{-2}$ in the observer frame~(Sherman et al., in prep.). The extragalactic DMs of both FRBs are likely dominated by the intracluster medium of the host galaxy clusters~(Connor et al., in prep.).


The PanSTARRS1~\citep[PS1;][]{2016arXiv161205560C} i-band images of galaxies coincident with the 90\% confidence localization region of these FRBs are displayed in Figure~\ref{fig:hosts_ppxf_fits}. We use \texttt{astropath} to calculate the association probability for each FRB to nearby galaxies~\citep{2021ApJ...911...95A}. The fields of both FRBs have been observed as part of the DESI Legacy Surveys in g, r, and z bands. For each FRB, we build a galaxy catalog by selecting resolved sources within 30\arcsec of the FRB with the \texttt{astro-datalab} Python library. To calculate an association probability, \texttt{astropath} requires the FRB position and error, as well as each galaxy’s position, magnitude (we use r-band), and half-light radius. We use the adopted priors recommended in \citet{2021ApJ...911...95A}, which assumes an exponential FRB angular offset distribution and an association probability that scales inversely to the number density of galaxies at a given magnitude (``exp'' and ``inverse'', respectively). We further assume a prior on an undetected host in flux-limited data, $P(U)=0.1$, which provides reliable and accurate estimates in realistic FRB host simulations \citep{2021arXiv211207639S}.
Following this procedure, we find that FRB 20220509G is associated to a host galaxy at (R.A. J2000, decl. J2000) = 18h50m41.92s, +70d14m33.95s with 1\% false association probability. This galaxy is cataloged as 2MASX J18504127+7014359 (J1850+70 hereafter) in the NASA Extragalactic Database~\citep{1991ASSL..171...89H}. FRB 20220914A is associated to a host galaxy at (R.A. J2000, decl. J2000) = 18h48m13.96s, +73d20m10.70s (J1848+73 hereafter) with 3\% false association probability. The analysis with $P(U)=0.5$ reveals a 6\% and 8\% false association probability for the likely hosts of FRB 20220509G and FRB 20220914A, respectively.


We obtained the optical spectrum of both the host galaxies with the Low-Resolution Imaging Spectrometer on the Keck I telescope~\citep[Keck-I/LRIS][]{1995PASP..107..375O}. We could only use the blue component of the detector due to instrument malfunction during the night of observations, so a mirror was used to direct light only into the blue arm. The light was dispersed using a 300/5000 grism. Single exposures of 1800~s and 500~s were obtained on 2022 October 18 using a 1\arcsecs slit at a position angle of 236.40\degr~and 299.95\degr~in good observing conditions with seeing of 0.84\arcsecs and 0.95\arcsecs for J1848+73 and J1850+70 respectively. The slit positions used to extract the galaxy spectra are indicated in Figure~\ref{fig:hosts_ppxf_fits}. The rest-frame line FWHM was approximately 9.5~$\AA$. The spectra were reduced with the standard \sw{lpipe} software~\citep{2019PASP..131h4503P} and calibrated using observations of the standard star BD+28 4211. We further scale the spectrum to match PS1 g-band photometry (described in \S~\ref{subsec:isophotal_analysis}) to account for slit losses.

The spectrum of J1848+73 exhibits strong emission lines and absorption features thus indicating a composition of young and old stellar populations in this galaxy~(Figure~\ref{fig:hosts_ppxf_fits}). We measure the spectroscopic redshift of the host galaxies using the Penalized PiXel-Fitting software~\citep[\sw{pPXF};][]{2017MNRAS.466..798C, 2022arXiv220814974C} by jointly fitting the stellar continuum and nebular emission using the MILES stellar library~\citep{2006MNRAS.371..703S}. The best \sw{pPXF} fit to the spectrum has a reduced-$\chi^2$ of 0.9076 (number of degrees of freedom, N $\sim$1000) and reveals a redshift of $0.1139 \pm 0.0001$. The Milky Way galactic dust extinction corrected measured line flux of [O II] and H$\beta$ lines are $(2.90 \pm 0.10) \times 10^{-16}~\mathrm{erg}~\mathrm{s}^{-1}\mathrm{cm}^{-2}$ and $(1.16 \pm 0.03) \times 10^{-17}~\mathrm{erg}~\mathrm{s}^{-1}\mathrm{cm}^{-2}$ respectively. The star formation rate (SFR) using the [O II] luminosity and calibrated using the \citet{1998ARA&A..36..189K} calibration is measured to be $0.14 \pm 0.10~\mathrm{M}_\odot \mathrm{yr}^{-1}$. We note that these SFR measurements are not corrected for the dust extinction within the host galaxy and hence, these SFRs serve as a lower limit on the true SFR.

The strong [Ca II], H$\beta$ and [Mg II] absorption features with [O II] emission are evident in the spectrum of J1850+70 thus indicating that it is an early-type galaxy~(Figure~\ref{fig:hosts_ppxf_fits}). The spectroscopic redshift of J1850+70 is also measured using \sw{pPXF}, where the best fit with a reduced-$\chi^2$ of 1.0166 (N $\sim$1000) indicates a redshift of $0.0894 \pm 0.0001$. The Milky Way galactic dust extinction corrected [O II] line flux is $(8.74 \pm 1.39) \times 10^{-17}~\mathrm{erg}~\mathrm{s}^{-1}\mathrm{cm}^{-2}$, which corresponds to an SFR of $0.04 \pm 0.01~\mathrm{M}_\odot \mathrm{yr}^{-1}$. An upper limit on the H$\beta$ line emission is $(2.18 \pm 1.04) \times 10^{-17}~\mathrm{erg}~\mathrm{s}^{-1}\mathrm{cm}^{-2}$. This corresponds to an [O II]/H$\beta$ $\gtrsim$ 2.71 at 1-sigma level, which is greater than the expected typical value for field galaxies~\citep{2004AJ....127.2002K}, thus indicating low star formation in this galaxy.


\section{Analysis Framework} \label{sec:analysis_framework}

In this section, we describe the analysis framework for deriving the properties of host galaxies using their photometric and spectroscopic data.

\subsection{Isophotal Analysis} \label{subsec:isophotal_analysis}

We executed photometry on archival images of PS1, Two Micron All Sky Survey~\citep[2MASS;][]{2006AJ....131.1163S} and ALLWISE~\citep{2014yCat.2328....0C} surveys. The 5-sigma limiting magnitude of 2MASS data for the J1848+73  galaxy are J = 19.7~mag, H = 18.8~mag and K$_{\mathrm{s}}$ = 18.1~mag. Due to shallow depth, this galaxy is marginally detected in 2MASS data, and hence, we do not include these data in our analysis. Furthermore, this galaxy is not detected in ALLWISE data. We iteratively fit elliptical isophotes to the PS1 i-band image of the galaxy using standard procedures defined in \sw{photutils}~\citep{larry_bradley_2022_6825092} to identify the isophote that captures $\gtrsim$95\% of the light from the galaxy. The best isophote indicated by our isophotal analysis has a semi-major axis of 4.644\arcsec~with an ellipticity of 0.326 (Figure~\ref{fig:hosts_ppxf_fits}). We convolve this aperture with the point spread function of all images to measure the instrumental magnitudes in all bands. This instrumental magnitude is then corrected using zero-point, interstellar dust reddening, and extinction to obtain the AB magnitudes~\citep{2018JOSS....3..695M, 1999PASP..111...63F}.

\begin{figure}
    \includegraphics[width=\columnwidth]{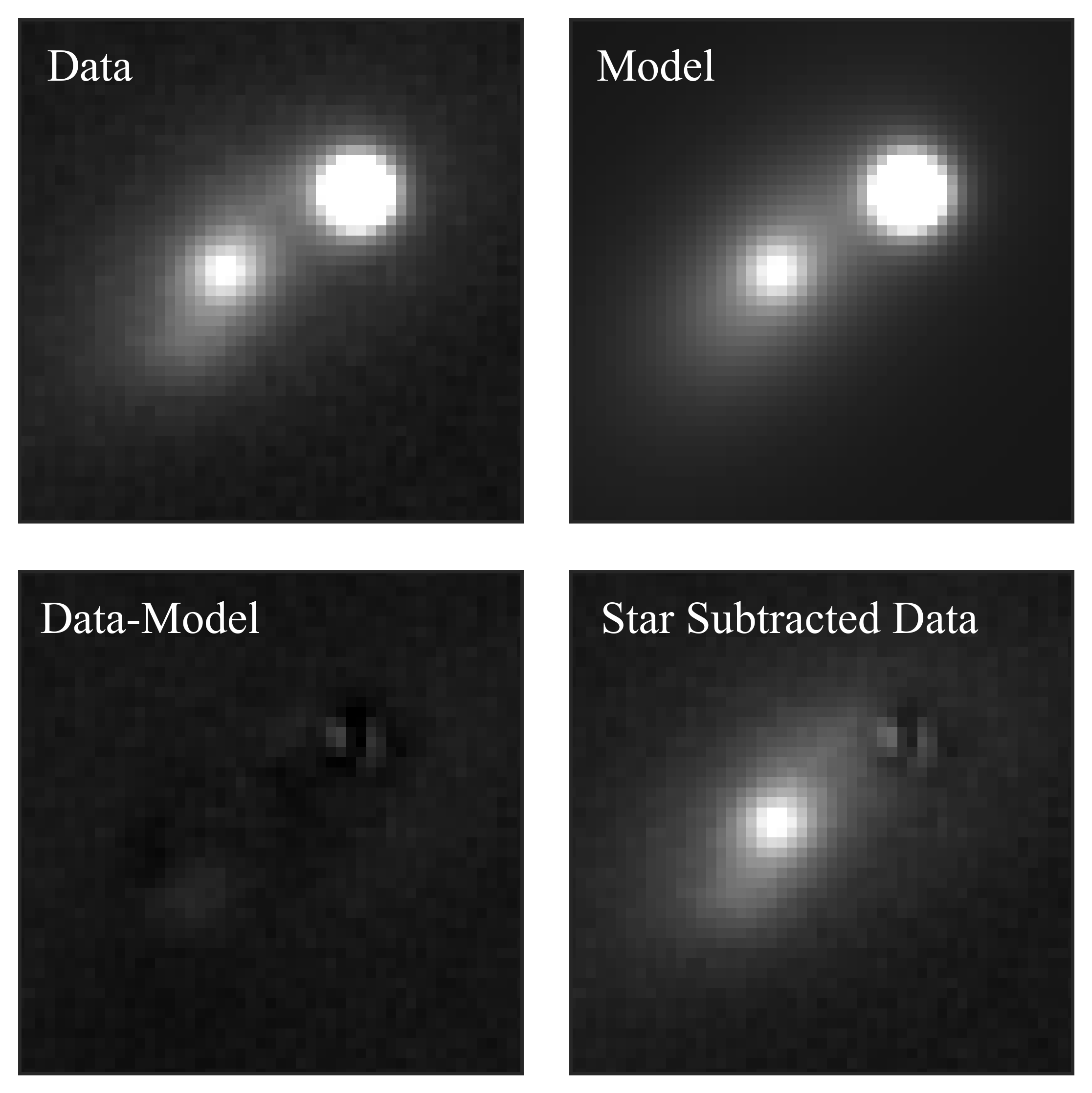}
    \caption{The host galaxy of FRB 20220509G has a bright star present at an angular separation of 4\arcsec, as can be seen in the i-band image of this galaxy in the top left panel. We model the star and the galaxy with circular and elliptical moffat profiles, respectively (top right panel), and test the quality of subtraction by subtracting this model from the data (bottom left panel). The total counts in a 5\arcsec aperture at the star's location are consistent with zero, thus confirming an acceptable subtraction quality. The final star-subtracted data is displayed in the bottom right panel.}
    \label{fig:subtraction_quality}
\end{figure}

The data for the J1850+70 galaxy are contaminated by the presence of a bright star at an angular separation of $\sim$4\arcsec. The typical method for photometry involves either masking the pixels at the location of the star or equivalently, using a smaller aperture focused at the center of the galaxy. However, we note that our galaxy is extended, and masking out those pixels will reduce its flux, and hence its stellar mass estimate. We confirm this by redoing our SED analysis (described in the following section) while using a smaller aperture size capturing the nuclear region of $\sim$2\arcsec radius. We note that while the recent SFR remains consistent with zero, the stellar mass ($\log{\mathrm{M}_\ast}$) drops by $\sim$5\%. Therefore, in order to perform photometry, we fit a circular moffat profile to the star and subtract it from our data. The quality of subtraction is assessed by jointly fitting an elliptical moffat profile to the galaxy and a circular moffat profile to the star and ensuring approximately zero counts in a 5\arcsecs aperture around the star (Figure~\ref{fig:subtraction_quality}). The pixel scale of 2MASS data is 1\arcsec/pixel with a typical FWHM of 2.5\arcsecs in all bands. Due to the compact point spread function of the star and very low counts of the galaxy, the star subtraction is poor in H and K$_{\mathrm{s}}$ bands. Hence, we do not include these two bands in our analysis. Furthermore, the signal-to-noise ratio of the galaxy detection is poor in ALLWISE W3 and W4 bands and hence they are not included in our analysis. 

The isophotal analysis of the star-subtracted i-band image of the galaxy reveals an elliptical flux-conservative profile with a semi-major axis of 15.48\arcsec~and an ellipticity of 0.52 (Figure~\ref{fig:hosts_ppxf_fits}). The axial ratio of its half-light isophote is 0.59. At low redshifts, the probability distribution of the axial ratio for spirals is flat whereas it rises for elliptical galaxies, thus indicating it is potentially an elliptical galaxy~\citep{2013MNRAS.434.2153R}. However, an axial ratio of 0.59 also implies a significant bulge dominance, which is typical of lenticular galaxies, and hence, this possibility cannot be ruled out based on the ellipticity measurements alone. In \S~\ref{subsec:sed_analysis}, we present more evidence to resolve the host galaxy classification for FRB 20220509G.

\begin{table}[ht!]
    \centering
    \begin{tabular}{lll}
        \toprule
        {Parameter} & {Value} & {Prior} \\
         \hline 
         SFH & & \\
         $\log{\mathrm{M}_\ast}~[\mathrm{M}_\odot]$ & 10 & Uniform(8, 12) \\
         $r_i$\footnote{Ratio of SFR in $i$th and its adjacent bin} & 0 & StudentT(0, 0.3, 2) \\
         $\log{z/z_\odot}$ & -0.2 & Uniform(-2.0, 0.2) \\
         N$_{\mathrm{bins}}$\footnote{Number of bins in the non-parametric star formation history} & 7 & -- \\
         $z_{\mathrm{red}}$ & $z^\prime$\footnote{Best-fit redshift from \sw{pPXF}} & Uniform($z^\prime$-0.01, $z^\prime$+0.01) \\
         \hline
         Dust Attenuation & & \\
         $\tau_{\mathrm{5500,~diffuse}}$\footnote{Opacity at 5500$\AA$ describing the attenuation of old stellar light} & 0.5 & $\mathcal{N}$(0.3, 1.0, 0, 4) \\
         \hline 
         Nebular Emission & & \\
         $U_{\mathrm{neb}}$\footnote{Nebular ionization parameter} & -2 & Uniform(-4, -1) \\
         $w_{\mathrm{eline}}$\footnote{Width of emission line amplitude prior \citep{2021ApJS..254...22J}} & 1 & Uniform(0.01, 100) \\
         $\sigma_{\mathrm{eline}}$\footnote{Emission lines broadening parameter} & 200 & Uniform(30, 500) \\
         \hline
         Dust Emission\footnote{Parameters from \citet{2007ApJ...657..810D} emission model} & & \\
         $U_{\mathrm{min,~dust}}$ & 1 & -- \\
         $Q_{\mathrm{PAH}}$ & 4 & -- \\
         $\gamma_{\mathrm{dust}}$ & 0.001 & -- \\
         \hline
         Spectral Calibration & & \\
         $\sigma_{\mathrm{smooth}}$\footnote{Spectral resolution} & 200 & Uniform(30, 500) \\
         \hline
    \end{tabular}
    \caption{Summary of free and fixed parameters used in our spectral energy distribution (SED) analysis.}
    \label{table:sed_params}
\end{table}

\begin{table}[ht!]
    \centering
    \begin{tabular}{lll}
        \toprule
        {Parameter} & {FRB 20220914A} & {FRB 20220509G} \\
         \hline
         $z_{\mathrm{red}}$ & 0.1139 $\pm$ 0.0001 & 0.0894 $\pm$ 0.0001 \\
        D$_{\mathrm{L}}$\footnote{Luminosity distance} [Mpc] & 534.94 & 412.95 \\
         $d$\footnote{Projected physical offset from galaxy center} [kpc] & 9.87 & 3.80 \\
         $r_e$\footnote{Effective radius} [kpc] & 2.67 & 6.64 \\
         $\log{\mathrm{M}_\ast}~[\mathrm{M}_\odot]$ & $9.99_{-0.09}^{+0.09}$ & $11.13_{-0.02}^{+0.02}$ \\
         SFR\footnote{Recent SFR averaged over the last 100~Myr} & $1.45_{-0.61}^{+1.05}$ & $0.08_{-0.04}^{+0.06}$ \\
         log(sSFR) [Gyr$^{-1}$] & $-0.82_{-0.22}^{+0.20}$ & $-3.23_{-0.29}^{+0.23}$ \\
         $\log{z/z_\odot}$ & $-0.92_{-0.03}^{+0.04}$ & $-0.11_{-0.03}^{+0.03}$ \\
         $A_V$ & $1.64_{-0.23}^{+0.22}$ & $0.19_{-0.04}^{+0.04}$ \\
         $U_{\mathrm{neb}}$ & $-3.28_{-0.03}^{+0.03}$ & $-3.53_{-0.27}^{+0.22}$ \\
         u $-$ r (Rest Frame) & 2.14$_{-0.11}^{+0.10}$ & 2.60$_{-0.02}^{+0.03}$ \\
         g $-$ r (Rest Frame) & 0.69$_{-0.04}^{+0.04}$ & 0.89$_{-0.01}^{+0.01}$ \\
         M$_{\mathrm{r}}$ (Rest Frame) & $-18.80_{-0.02}^{+0.02}$ & $-21.38_{-0.01}^{+0.01}$ \\
         Milky Way E(B - V) & $-0.36$ & $-0.06$ \\
         \hline
    \end{tabular}
    \caption{Summary of observed and derived parameters of the host galaxies of the two FRBs presented in this article. Note that all the derived galaxy properties have been measured using the SED analysis. The quoted measurements are the 16th, 50th and 84th percentiles.}
    \label{table:sed_results}
\end{table}

\begin{figure*}
    \includegraphics[width=\textwidth]{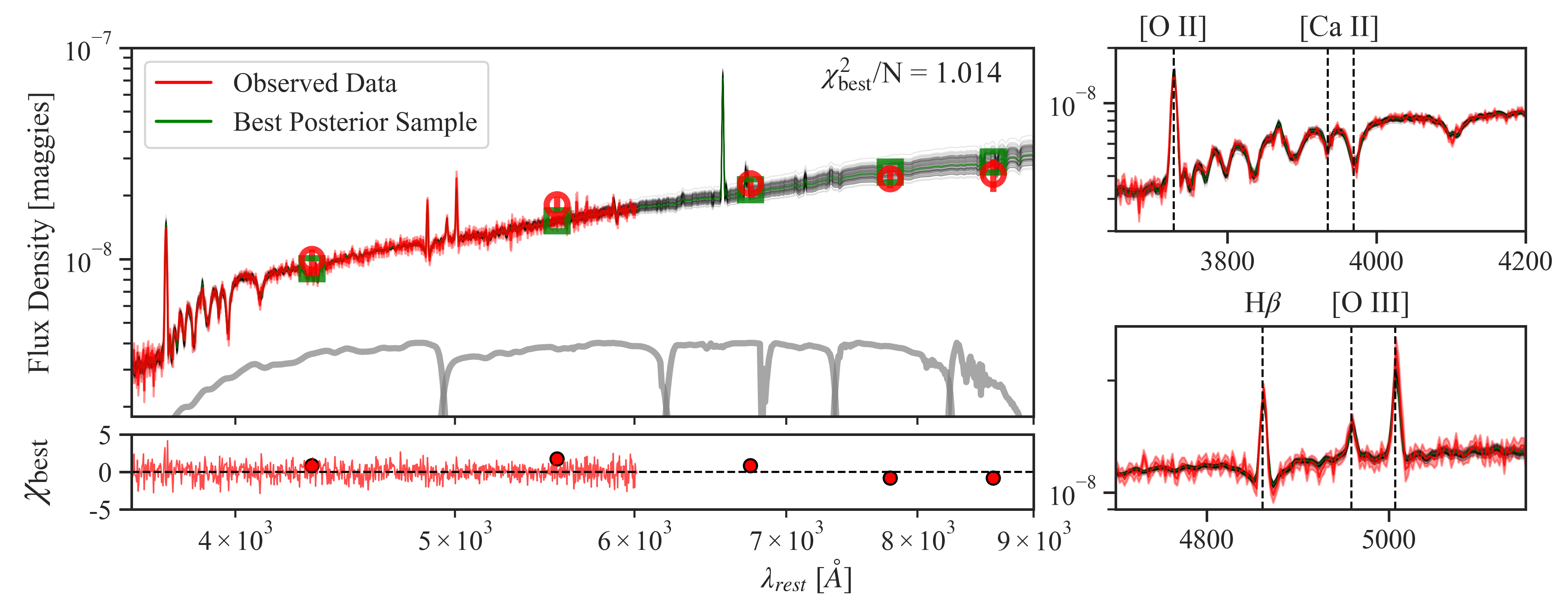}
    \includegraphics[width=\textwidth]{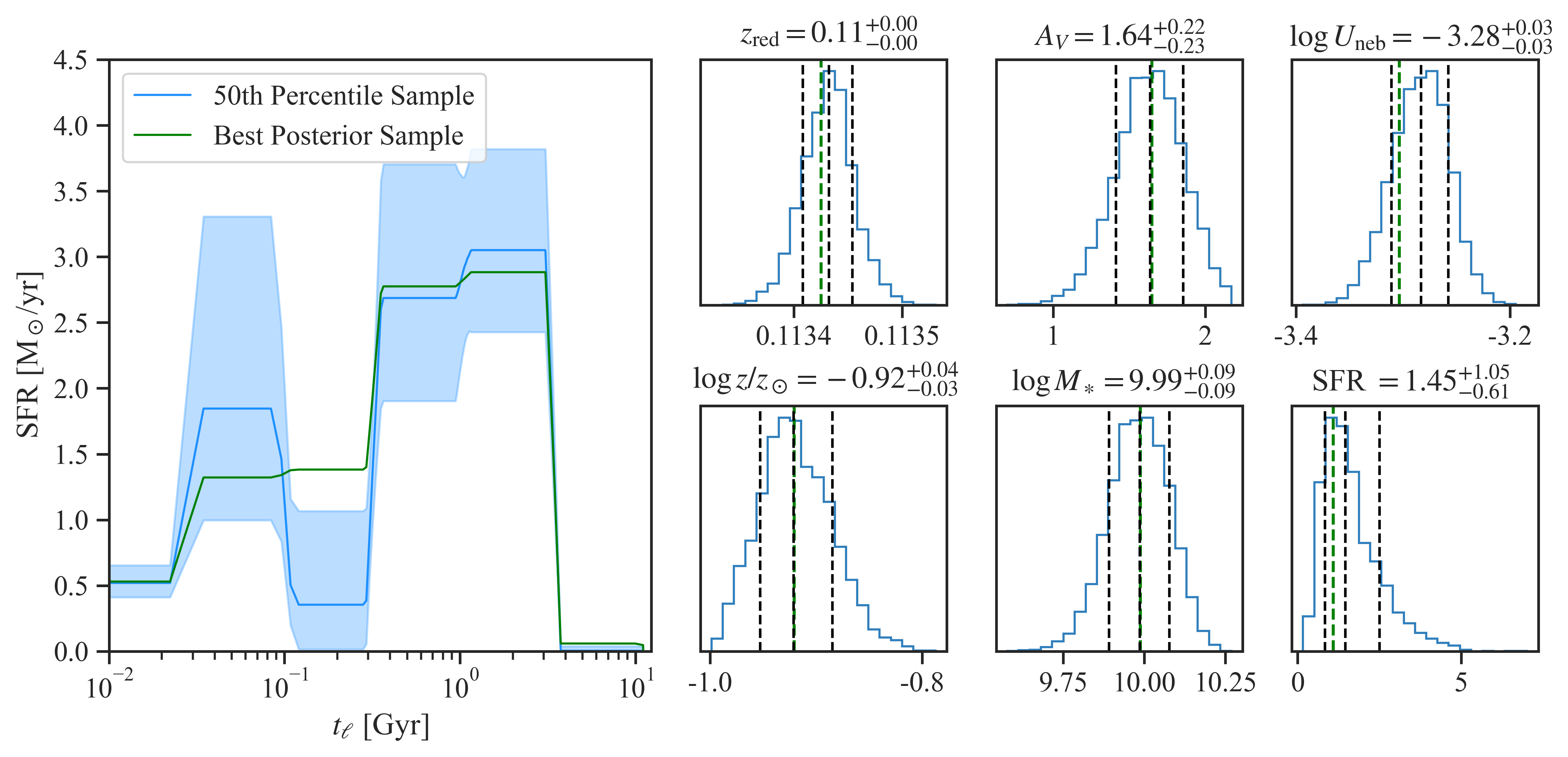}
    \caption{The results from a non-parametric star formation history SED fit to observed spectroscopy and photometry data (red) of the host galaxy of FRB 20220914A, J1848+73. The top panel shows the SED fit with residuals for the best posterior sample (green) and 100 random posterior samples (black). The plots zoomed at various absorption and emission features indicate the accuracy of our SED fits. The bottom panel displays the star formation history and constrained galaxy parameters with 16th and 86th percentiles indicated. The derived parameters for J1848+73 are consistent with a typical star-forming galaxy.}
    \label{fig:sed_fit_elektra}
\end{figure*}

\subsection{SED Analysis} \label{subsec:sed_analysis}

We use the stellar population synthesis modeling software \sw{Prospector}~\citep{2021ApJS..254...22J} which uses the Flexible Stellar Population Synthesis~\citep[\sw{FSPS};][]{2009ApJ...699..486C, 2010ascl.soft10043C}, to determine the stellar properties of our host galaxies. We simultaneously model and fit for the observed photometry and spectroscopy. Due to underestimated photometric errors and imperfect subtraction of the star, we assume additional 10\% photometric errors for both the galaxies. We initialize the redshift to the value obtained from \sw{pPXF} fits with a uniform prior width of 1\%.  We chose to use a continuity non-parametric star formation history with 7 bins to avoid systematics induced by parametric star formation histories ~\citep{2013ARA&A..51..393C, 2017ApJ...837..170L}. We assume the \citet{2001MNRAS.322..231K} initial mass function. We include nebular continuum and line emission in our model, which is based on the \sw{CLOUDY} implementation within \sw{FSPS}~\citep{2013RMxAA..49..137F}. We tie the nebular emission metallicity to the stellar metallicity and float the nebular ionization parameter. The nebular emission model assumes that all of the nebular emission is produced by the young stellar population, which may not always be true in galaxies where they are instead powered by active galactic nuclei or shocks~\citep{2006ApJ...648..281Y}. To account for such cases, we marginalize the amplitude of emission lines in our observed spectrum. We include dust emission in the model but fix all the dust emission parameters due to lack of good quality data at infrared wavelengths~\citep{2007ApJ...657..810D}. We use spectral smoothing and a 12th-order Chebyshev polynomial for parameterized spectrophotometric calibration. The set of parameters in our model and corresponding priors are summarized in Table~\ref{table:sed_params}. We sample from the posterior using the ensemble sampler \sw{emcee}~\citep{2013PASP..125..306F}. For a discussion on best practices in SED modeling, we refer the reader to the appendix and the references therein.

\begin{figure*}
    \includegraphics[width=\textwidth]{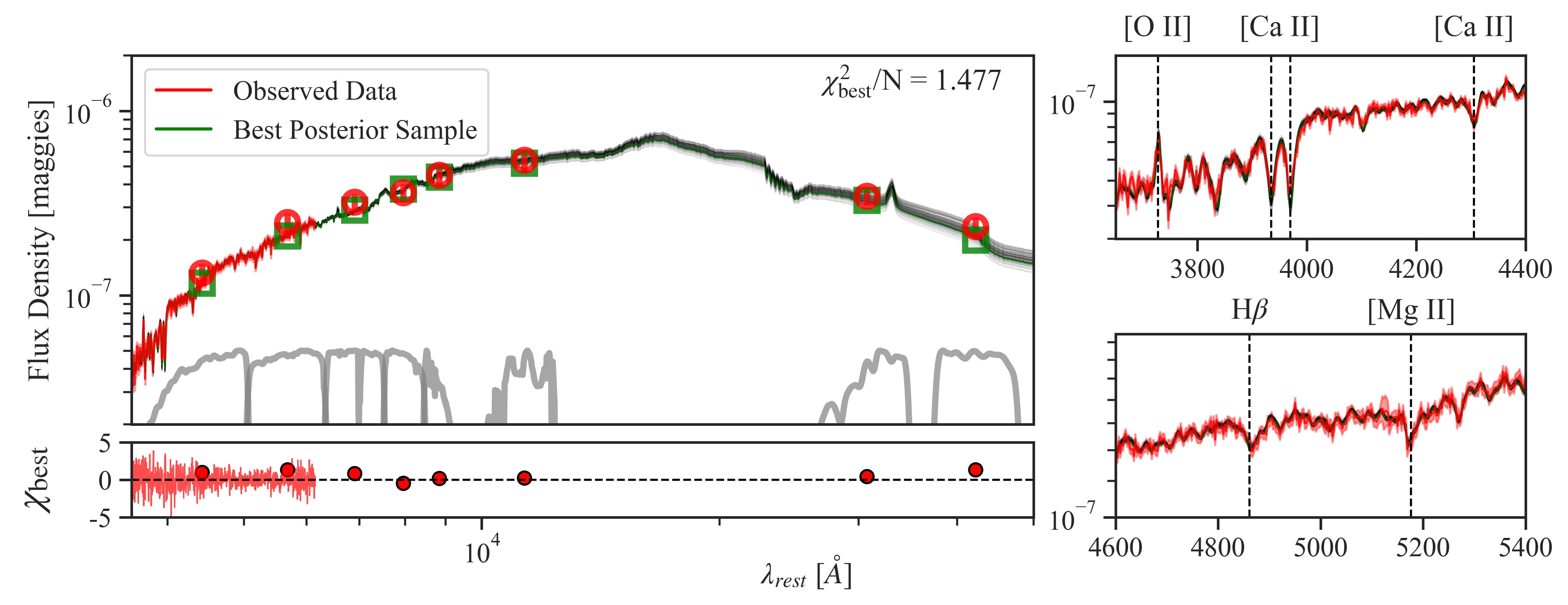}
    \includegraphics[width=\textwidth]{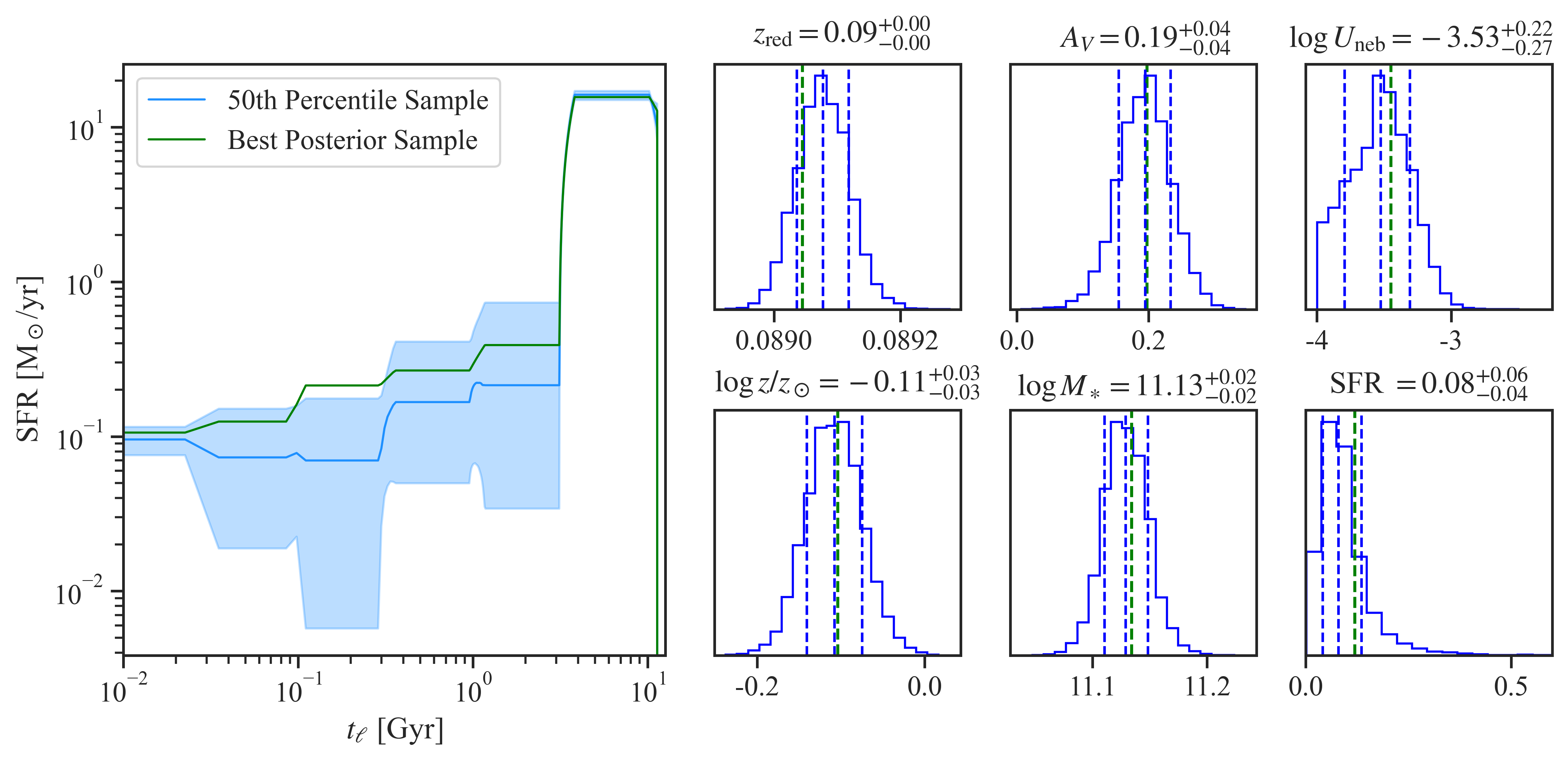}
    \caption{The results from a non-parametric star formation history SED fit to observed spectroscopy and photometry data (red) of the host galaxy of FRB 20220509G, J1850+70. The top panel shows the SED fit with residuals for the best posterior sample, along with 100 random posterior samples (black). The zoomed version of the plots display the accuracy of our fits to various absorption and emission features. The bottom panel displays the star formation history and constrained galaxy parameters with 16th and 86th percentiles indicated. The high stellar mass and low recent SFR of J1850+70 with low dust attenuation are indicative of a massive elliptical galaxy.}
    \label{fig:sed_fit_jackie}
\end{figure*}

The SED fits for the host galaxies of FRB 20220914A and FRB 20220509G are displayed in Figure~\ref{fig:sed_fit_elektra} and \ref{fig:sed_fit_jackie} respectively and the corresponding observed and derived parameters are summarized in Table~\ref{table:sed_results}. We observe that all the nebular emission and absorption features, along with the photometry, are well fit by the model with a reduced-$\chi^2$ of 1.014 and 1.477 (N $\sim$1000) for the two galaxies. The star formation history of the host galaxy of FRB 20220914A indicates a variety of stellar population ages, which is consistent with our inference from the spectrum, as discussed in \S~\ref{sec:observations}. The high dust attenuation, $A_V = 1.64_{-0.23}^{+0.22}$ and stellar mass, $\log{\mathrm{M}}_\ast (\mathrm{M}_\odot) = 9.99_{-0.09}^{+0.09}$ with significant ongoing star formation averaged over the last 100~Myr, SFR = $1.45_{-0.61}^{+1.05}~{\mathrm{M}}_\odot$/yr indicates that it is a Milky Way-like star-forming spiral galaxy. On the other hand, the star formation history of FRB 20220509G indicates a very old stellar population, which is also consistent with it being an early-type galaxy as discussed in \S~\ref{sec:observations}. The low dust attenuation $A_V = 0.19_{-0.04}^{+0.04}$ supports our argument of it being an elliptical galaxy, as discussed in \S~\ref{subsec:isophotal_analysis}. Further, a high stellar mass, $\log{\mathrm{M}}_\ast (\mathrm{M}_\odot) = 11.13_{-0.02}^{+0.02}$ and a consistent with zero SFR averaged over the last 100~Myr, SFR = $0.08_{-0.04}^{+0.06}~{\mathrm{M}}_\odot$/yr implies that this is a quiescent galaxy. We note that SFR measured using [O II] emission line luminosity in 
\S~\ref{sec:observations} is a lower limit on the actual SFR since they are not corrected for dust attenuation within the host galaxy itself. The SFRs measured from our SED analysis are corrected for the dust attenuation within the host galaxies and hence, are consistent with the lower limits on SFRs as reported in \S~\ref{sec:observations}.

\begin{figure*}
    \includegraphics[width=\textwidth]{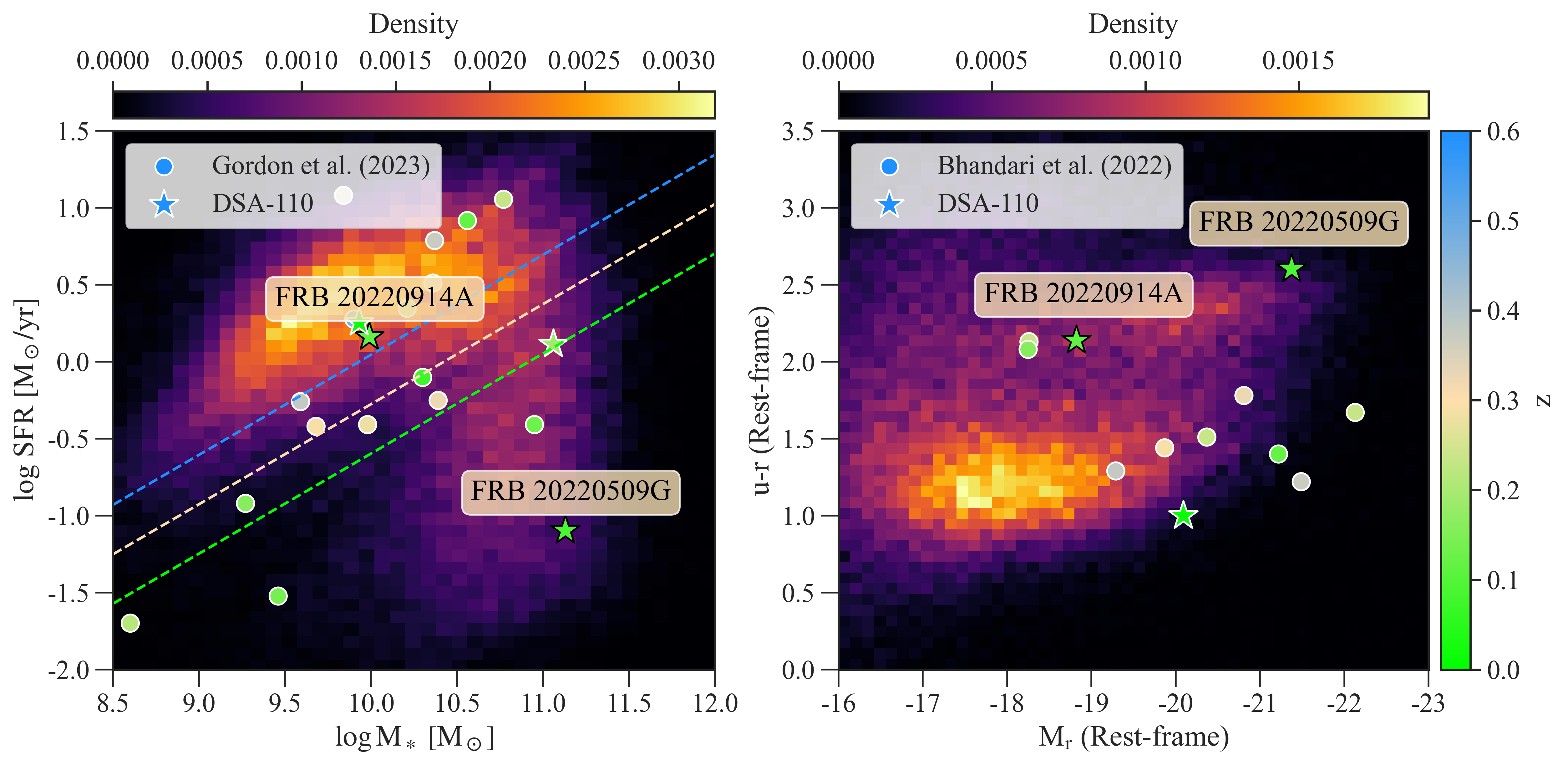}
    \caption{Comparison of the host galaxies of FRBs (including the sample published in \citet{2023arXiv230205465G}) with the background galaxies from the PRIMUS dataset~\citep{2013ApJ...767...50M}. The stellar mass-SFR plot in the left panel with redshift evolution of the boundary between star-forming and quiescent galaxies~\citep{2013ApJ...767...50M} indicates that while the host galaxy of FRB 20220914A is a typical star-forming galaxy, the host galaxy of FRB 20220509G stands out as a quiescent galaxy in the population of known FRB hosts. The color-magnitude diagram in the right panel shows that the host galaxy of FRB 20220509G is red and composed of older stellar population as compared to the rest of the FRB hosts population~\citep{2022AJ....163...69B}.}
    \label{fig:gal_sfs_cmd}
\end{figure*}

\section{Nature of FRB Progenitors} \label{sec:frb_progenitors}

In this section, we compare the host galaxies of FRB 20220509G and FRB 20220914A with the hosts of other FRBs, the background galaxy population, and the hosts of other transient populations. Along with our two FRBs, we include the sample of 17 non-repeating FRBs published in \citet{2023arXiv230205465G}, which includes refined host properties computed with non-parametric SED modeling for FRBs published in \citet{2022AJ....163...69B}, \citet{2022arXiv221116790B}, \citet{2020ApJ...903..152H} and \citet{2021ApJ...917...75M}. We also include the previously reported non-repeating FRBs discovered by the DSA program, namely FRB 190523~\citep{2019Natur.572..352R}, and FRB 20220319D~\citep{2023arXiv230101000R} in our comparison sample. We then attempt to demonstrate the formulation of delay-time distribution analysis for FRB progenitors using the two FRBs reported in this article.

\subsection{Comparison with Background Galaxy Population} \label{subsec:field_galaxies}

We use the GALEX-SDSS-WISE Legacy Catalog~\citep[GSWLC;][]{2018ApJ...859...11S} for background galaxies population with redshift $\leq 0.2$ and PRIMUS~\citep{2013ApJ...767...50M} dataset for background galaxies population with redshift $\geq 0.2$ but $\leq 0.6$ to match the characteristic redshift range of FRBs. Therefore, our background galaxies population dataset comprises $\sim$77,000 galaxies with an approximately uniform distribution of galaxy redshifts. We note that there are significant systematics involved in such comparative analysis. These systematics arise from the differences in the SED-modeling approaches, such as parameterization of the star formation histories and measurements of recent SFR. For a more accurate comparison, one must use derived galaxy properties with non-parametric star formation history SED modeling. However, due to unavailability of such public dataset, we resort to using parametric derived background galaxy properties.

The left panel of Figure~\ref{fig:gal_sfs_cmd} shows the distribution of FRB hosts in the space of stellar mass and recent SFR along with the redshift evolution of the boundary between star-forming and quiescent galaxies~\citep{2013ApJ...767...50M}. We observe that the host of FRB 20220914A is a typical star-forming galaxy. While most of the FRB hosts lie around the star-forming main sequence, the host of FRB 20220509G is exceptional as a quiescent galaxy. Recently, \citet{2023arXiv230205465G} used the mass-doubling number criterion of \citet{2022ApJ...926..134T} to classify galaxies as star-forming, transitioning and quiescent. Since this criterion was developed on galaxy properties derived using non-parametric star formation histories, it is more appropriate to classify the hosts of our two FRBs using the mass-doubling number. The mass-doubling number for the hosts of FRB 20220914A and FRB 20220509G are 1.823 and 0.007, thus classifying them as star-forming and quiescent galaxies respectively. This is consistent with our previous arguments.

The right panel of Figure~\ref{fig:gal_sfs_cmd} shows the color-magnitude diagram with the distribution of background galaxies and FRB hosts plotted. Due to the unavailability of colors and magnitudes of the 23 FRB hosts published in \citet{2023arXiv230205465G}, we use the data from \citet{2022AJ....163...69B}. While most of the FRB hosts are late-type galaxies with young stellar populations and significant ongoing star formation, the host of FRB 20220509G stands out as an early-type galaxy with an old stellar population in the red cloud of the background galaxies population in the color-magnitude diagram. 

We further compare the stellar mass and SFR of the host of FRB 20220509G with the typical values for elliptical and spiral galaxies, computed using the galaxy classifications in Galaxy Zoo dataset~\citep{2011MNRAS.410..166L}. We note that the typical redshift range for galaxies in the Galaxy Zoo dataset is $\lesssim$0.2, which is consistent with the redshift of the host of FRB 20220509G. All the queries were performed using \sw{CasJobs}~\citep{2005cs........2072O}. The stellar mass and SFRs for typical spiral galaxies are $\log{\mathrm{M}}_\ast (\mathrm{M}_\odot) = 10.77_{-0.62}^{+0.39}$ and $\log{\mathrm{SFR}}~({\mathrm{M}}_\odot$/yr) = $0.44_{-0.74}^{+0.46}$ whereas for typical elliptical galaxies, $\log{\mathrm{M}}_\ast (\mathrm{M}_\odot) = 11.24_{-0.56}^{+0.36}$ and $\log{\mathrm{SFR}}~({\mathrm{M}}_\odot$/yr) = $-0.97_{-0.58}^{+1.11}$. Both the stellar mass and SFR for the host galaxy of FRB 20220509G are consistent with elliptical galaxies, thus providing additional evidence for it being an elliptical galaxy.

\subsection{Comparison with Extragalactic Transients} \label{subsec:compare_with_other_transients}

\begin{figure}
    \includegraphics[width=\columnwidth]{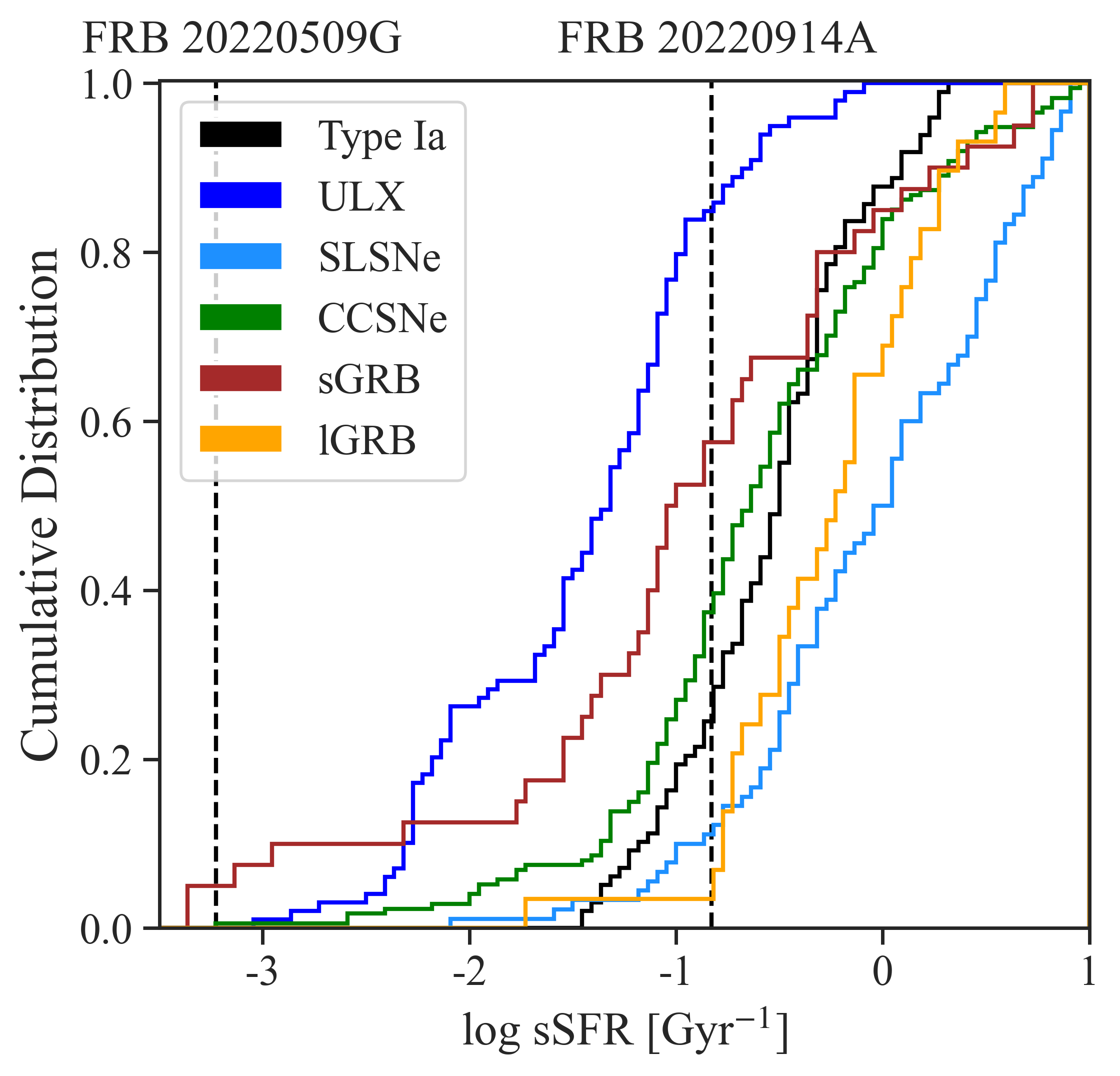}
    \caption{The cumulative distributions of sSFRs for different extragalactic transients, evaluated at redshift z=0 (see text for details). The vertical lines show the sSFRs of the new FRB hosts we present herein.}
    \label{fig:compare_other_transients}
\end{figure}

A comparison of the environment of transients is important to identify the possible similarities in their progenitors and formation channels. Several such investigations have been conducted in the past~\citep{2022AJ....163...69B, 2021ApJ...907L..31B, 2020ApJ...895L..37B, 2020ApJ...903..152H, 2020ApJ...899L...6L, 2021ApJ...917...75M, 2020ApJ...905L..30S}. We augment these works by specifically considering our two new FRB hosts. We compare the specific SFRs (sSFRs) of the host galaxies of FRBs with Type Ia supernovae~\citep{2010ApJ...722..566L}, ultra-luminous X-ray sources~\citep[ULX;][]{2020MNRAS.498.4790K}, super luminous supernovae~\citep[SLSNe;][]{2021ApJS..255...29S, 2021MNRAS.503.3931T}, core-collapse supernovae~\citep[CCSNe;][]{2021ApJS..255...29S, 2021MNRAS.503.3931T}, short gamma-ray bursts~\citep[sGRB;][]{2022ApJ...940...57N} and long gamma-ray bursts~\citep[lGRB;][]{2015A&A...581A.102V, 2021MNRAS.503.3931T}. We note that the hosts of transients in our comparison sample have a huge variance. The redshifts of FRB hosts in our sample are $z = 0.214_{-0.120}^{+0.125}$, where quoted values are the 16th, 50th and 84th percentiles of the redshift distribution. On the other hand, the redshift distribution of other transient's hosts is quite varied, including $z = 0.009_{-0.006}^{+0.018}$ for ULX, 
$z = 0.226_{-0.114}^{+0.106}$ for SLSNe, 
$z = 0.039_{-0.022}^{+0.047}$ for CCSNe, 
$z = 0.485_{-0.262}^{+0.320}$ for sGRB and $z = 0.283_{-0.194}^{+0.422}$ for lGRB. Therefore, for a fair comparison, one must account for the redshift evolution of the galaxy star-forming main sequence. The stellar mass and SFR need to be corrected to statistically represent all the galaxies at the present epoch. To this end, we adopt the formulation developed by \citet{2021ApJ...907L..31B} to convert the stellar mass and SFRs of all hosts of transients to their respective values at z = 0, where the $p$-value of stellar mass and SFR relative to the distribution of star-forming galaxies is conserved at the redshift of the galaxy and the current epoch.


Figure~\ref{fig:compare_other_transients} shows the cumulative distributions of sSFR in the hosts of the different transient samples, together with the hosts of FRBs 20220914A and 20220509G. The sSFR of the FRB 20220914A host is consistent with essentially all transient populations. However, only sGRBs have been observed (among the samples under consideration) in galaxies with a similarly low sSFR as the host of FRB 20220509G. This is consistent with a scenario wherein, like sGRBs, FRB 20220509G may have occurred long after the star-formation event that formed its progenitor \citep[e.g.,][]{2022ApJ...940L..18Z}. \citet{2014MNRAS.441.2433R} also highlighted the possibility of FRB progenitor formation in binary neutron star mergers, which give rise to sGRBs. As above, similar results are obtained for stellar mass and SFR distributions.

We choose not to quantitatively compare the distributions of host-galaxy properties of these transient samples and the FRB host population discussed above. Optical host selection effects, where the magnitude-limited data may lead to misidentification of the host galaxies and only brighter hosts are chosen for further analysis, affect the stellar mass and SFR distributions, increasing the median values of respective parameters~\citep{2021arXiv211207639S}. The inconsistency in the SED-analysis approaches and recent SFR indicators used for deriving the galaxy properties of all transients introduces systematics that are difficult to quantify. For example, \citet{2021MNRAS.503.3931T} use a parametric exponentially declining star-formation history model to derive present-day star-formation rates for the CCSNe, SLSNe, and lGRBs included in Figure~\ref{fig:compare_other_transients}, whereas we use a non-parametric star-formation history. A more detailed analysis addressing some of these issues will be presented in a future work with a bigger FRB sample (Law et al., in prep.).


\begin{figure*}
    \includegraphics[width=\textwidth]{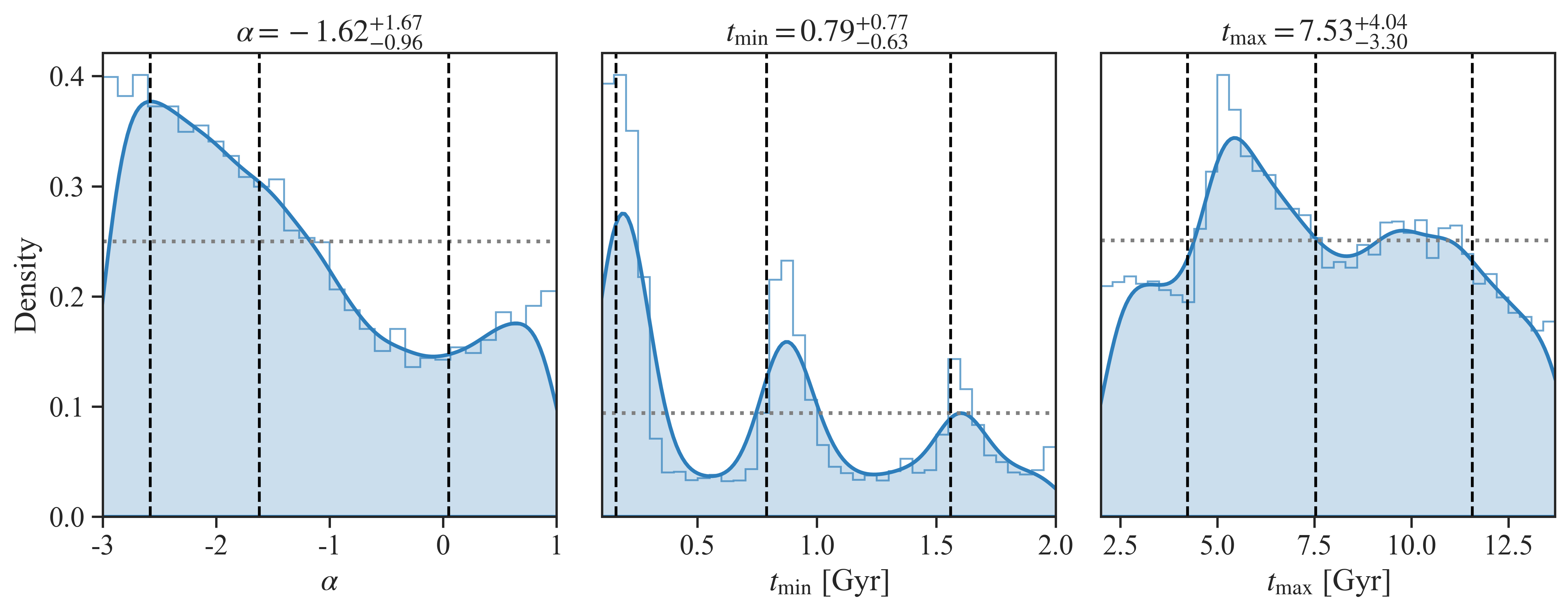}
    \caption{The kernel density estimates of the posteriors of the delay-time distribution parameters for FRBs constrained using non-parametric star formation histories of the host galaxies of FRB 20220914A and FRB 20220509G. The gray dotted line depicts our priors on these parameters and vertical lines show the 16\%, 50\% and 84\% credible regions. The multi-peaked feature in $t_{\mathrm{min}}$ posterior distribution is a characteristic feature embedded from non-parametric star formation histories.}
    \label{fig:delay_time_distribution}
\end{figure*}



\subsection{Delay-Time Distribution} \label{subsec:dtd}

Analyses of the host galaxies of transients yield information of the underlying stellar populations, which can allow us to put novel constraints on their progenitor channels. The delay-time distribution of transients can give us insights into the birth properties of their progenitors, as well as assist in disentangling multiple progenitor possibilities. For example, models of single-degenerate ONe/CO white dwarf --helium star binary channel of accretion-induced collapse events, which lead to the formation of intermediate-mass binary pulsars with short orbital periods, predict short delay times~\citep{2020RAA....20..135W}. On the other hand, models of single-degenerate ONe white dwarf -- red giant binary channel of accretion-induced collapse events, which lead to the formation of young millisecond pulsars in globular clusters, predict longer delay times~\citep{2020RAA....20..135W}. Furthermore, the delay-time distribution of transients is also a valuable probe of their formation rates. The expected local binary neutron star merger rate evolution computed using the delay-time distribution of sGRBs has been found to be consistent with constraints from gravitational wave observations~\citep{2022ApJ...940L..18Z}. This also affirmed these binary compact object mergers as the progenitors of sGRBs~\citep{2022ApJ...940L..18Z}.

Motivated by such studies, we attempt to constrain the delay-time distribution of FRBs using our two galaxy cluster FRBs. We note that the delay-time distributions for repeating and non-repeating FRBs may be different due to possible differences in their progenitor channels. Here, we focus on computing the delay-time distribution using our two apparently non-repeating FRBs. We define the delay time, $t_d$, as the time between the recent star burst in the galaxy and the time an FRB occurs. Essentially, $t_d = t_\ast + t_{\mathrm{age}}$, where $t_\ast$ is the time between the formation of progenitor stars and the formation of the FRB progenitors, and $t_{\mathrm{age}}$ is the age of the FRB source. In this initial analysis, we assume that $t_{\mathrm{age}} \ll t_\ast$, and hence, $t_d \approx t_\ast$. Following the formulation described in \citet{2022ApJ...940L..18Z}, we parameterize the delay-time distribution as a power law distribution in the range of $t_{\mathrm{min}}$ to $t_{\mathrm{max}}$ of stellar evolution timescale with a spectral index $\alpha$,

\begin{equation}
p(t_d,~\alpha,~t_{\mathrm{min}},~t_{\mathrm{max}}) = \begin{cases} \mathcal{N}t_d^\alpha, & t_{\mathrm{min}} \leq t_d \leq t_{\mathrm{max}} \\ 0, & \mathrm{otherwise} \end{cases},
\end{equation}

where $\mathcal{N}$ is the normalization. For a given host galaxy $i$ and a star formation history posterior sample $j$, the expected rate of FRBs $\dot{n}_i^j$ at redshift $z_i^j$ is defined as,

\begin{equation}
    \dot{n}_i^j = \int_{z^\prime = \infty}^{z^\prime = z_i^j} p(t_{\ell b}^\prime - t_{\ell b} | \alpha,~t_{\mathrm{min}},~t_{\mathrm{max}}) \lambda \psi_i^j(z^\prime) \frac{dt}{dz}(z^\prime)dz^\prime,
\end{equation}

where $t_{\ell b}^\prime$ and $t_{\ell b}$ are the lookback times at redshifts $z^\prime$ and $z_i^j$ respectively, $\lambda$ is the FRB source formation efficiency, which has been assumed to be $10^{-5}~\mathrm{M}_\odot^{-1}$, $\psi_i^j(z^\prime)$ is the non-parametric star formation history derived using \sw{Prospector} and $dt/dz$ is defined using the standard cosmological model. Assuming that the probability of occurrence of an FRB follows a Poisson distribution, the hyperlikelihood of observing the FRB from the particular galaxy can be written as,

\begin{equation}
    \mathcal{L}(\psi_i|\alpha,~t_{\mathrm{min}},~t_{\mathrm{max}}) \approx \frac{\mathcal{A}}{N} \sum_{j=1}^{N} (\dot{n}_i^j \Delta t) e^{-\dot{n}_i^j \Delta t},
\end{equation}

where $\Delta t$ is a fiducial observing time of 10~yr and $\mathcal{A}$ is the normalization. Assuming that our observations of FRB 20220914A and FRB 20220509G are independent, the hyperposterior is,

\begin{equation}
    P(\alpha,~t_{\mathrm{min}},~t_{\mathrm{max}} | \mathrm{obs}) \propto \Pi_{i = 1}^{2} \mathcal{L}(\psi_i|\alpha,~t_{\mathrm{min}},~t_{\mathrm{max}}) \times \pi,
\end{equation}

where $\pi(\alpha,~t_{\mathrm{min}},~t_{\mathrm{max}})$ is the prior on the delay-time distribution parameters, which are uniform in the range [-3, 1], [1~Myr, 2~Gyr] and [2~Gyr, 13.7~Gyr] respectively. We use precomputed grids of likelihoods and interpolate when evaluating the likelihood function. We use the \sw{dynesty} nested sampler~\citep{2020MNRAS.493.3132S} in the framework of \sw{Bilby}~\citep{2019ApJS..241...27A} for generating posterior distributions.

Our constraints on the delay-time distribution parameters are shown in Figure~\ref{fig:delay_time_distribution}. Given the small sample size, we cannot make meaningful statements regarding the posteriors of the three delay-time distribution parameters. Nevertheless, it is evident that the three parameters are not correlated. The multiple peaks in the posterior distribution of $t_{\mathrm{min}}$ indicate the importance of non-parametric star formation histories in constraining the delay-time distribution parameters, since all possible star bursts are taken into account, which would otherwise be missed in a parametric star formation history. Future studies with a bigger hosts sample can help to constrain these parameters better and shed some light on the evolutionary histories of FRB sources, and the FRB rate evolution with redshift.

\section{Summary and discussion} \label{sec:discussion}

FRBs have been found in a wide variety of environments~\citep{2022A&ARv..30....2P}, including star-forming regions in dwarf galaxies~\citep{2017ApJ...843L...8B, 2022arXiv221116790B}, spiral galaxies~\citep{2021ApJ...917...75M, 2020Natur.577..190M, 2021ApJ...919L..23F}, at significant offsets from star-forming regions~\citep{2021ApJ...908L..12T} and globular clusters~\citep{2022Natur.602..585K, 2021ApJ...910L..18B}. However, none have been previously associated with galaxy clusters. This and our companion paper, Connor et al.~(in prep.), report the discovery of two FRBs in massive galaxy clusters. The host galaxy of FRB 20220914A resides in the galaxy cluster Abell 2311~\citep{1958ApJS....3..211A} with M$_{180} = 2.4 \times 10^{14}~\mathrm{M}_\odot$ as per the DESI Legacy Imaging Surveys Data Release 9 (DR9) group/cluster catalog~\citep{2019AJ....157..168D} and the host galaxy of FRB 20220509G resides in the galaxy cluster Abell 2310~\citep{1958ApJS....3..211A} with M$_{180} = 2.5 \times 10^{14}~\mathrm{M}_\odot$. Out of the 21 FRBs that we consider as a sample (see \S~\ref{sec:frb_progenitors}), only $\sim$9.5\% of the FRBs are found in galaxy cluster environments. This is broadly consistent with the value of $\sim$10\% for the overall fraction of stellar mass in galaxy clusters~\citep{1998ApJ...503..518F}. However, this result may be surprising if the occurrence of FRBs is driven by ongoing star formation, as galaxy clusters contribute negligibly to the present-day cosmic star formation rate density \citep[e.g.,][]{2017ApJ...844L..23C}.

The SFR of galaxies is very well known to be correlated with the galaxy number density \citep{2004MNRAS.353..713K}. The galaxies at the core of the galaxy clusters have lower SFRs as compared to the infalling galaxies~\citep{2018ApJ...857...71B}. The recent SFR of the host galaxy of FRB 20220914A, which is a typical star-forming galaxy, is marginally higher than the typical SFR of galaxies in clusters at a cluster-centric distance of R/R$_{200} \sim$ 0.46~\citep{2016ApJ...816L..25P}. Given that the galaxy clusters are extremely effective at cutting off star formation in galaxies by stripping off the cold gas needed for stellar birth, significant star formation in a galaxy close to the core of the cluster is unusual. On the other hand, the host galaxy of FRB 20220509G is a red, old, massive elliptical galaxy, with low SFR, which is typical of quenched galaxies found in galaxy clusters~\citep{2018MNRAS.475..523L}. Notably, this is the first example of a likely massive elliptical FRB host galaxy.

The discovery of FRBs in spiral arms of late-type galaxies and galaxies with higher sSFR supports that FRBs should have short delay times. Although, while most of the FRBs found to date are associated with star-forming galaxies, the quiescent, elliptical host of FRB 20220509G adds diversity to the known FRBs host galaxy population. The origin of FRBs in quiescent elliptical galaxies and globular clusters adds to the evidence that some FRB progenitors have longer delay times. Together, these environments are inconsistent with a single population, thus hinting towards a broad delay-time distribution and suggesting multiple formation channels for FRBs. The origin of FRB 20220509G in an old stellar population disfavors the possibility of formation by young highly magnetized magnetars in a core-collapse supernova. This is further supported by the fact that only 0.3\% of the core-collapse supernovae occur in elliptical galaxies~\citep{2022ApJ...927...10I}.

The old stellar population in elliptical galaxies supports multiple possibilities about the progenitor of FRB 20220509G. The likelihood of the formation of binary neutron stars in old elliptical galaxies with negligible ongoing star formation opens up the possibility of an FRB source formed via binary neutron star merger~\citep{2022MNRAS.512.2654P, 2018A&A...615A..91B, 1989Natur.340..126E, 1992ApJ...395L..83N}. Secondly, this particular host environment also supports progenitor formation channels in globular cluster environments due to their higher number density in elliptical galaxies~\citep{2020NatAs...4..153L}. The high mass of the host galaxy could also favor an accretion-induced collapse of the white dwarf to neutron star~\citep{2019Natur.572..352R}. The remnant white dwarf formed in a typical binary white dwarf merger has been long known as a probable progenitor of Type Ia supernovae, where 99\% of Type Ia supernovae in elliptical galaxies likely occur via this formation channel~\citep{2011NewA...16..250L}. If one of the merging white dwarfs has a significant magnetic field, the merger may result in the formation of a magnetar, which can then power an FRB~\citep{2001MNRAS.320L..45K, 2013ApJ...776L..39K, 2020MNRAS.492.3753K}. Similar formation channels were also proposed by \citet{2022Natur.602..585K} upon the association of FRB 20200120E with a globular cluster in M81 due to the high probability of formation of binaries with short orbital periods in globular clusters~\citep{2020RAA....20..135W, 2013A&A...558A..39T}. The horizon of research in modeling the progenitors of FRBs must be broadened to incorporate such formation channels of these exotic transients.

\begin{acknowledgments}

The authors thank staff members of the Owens Valley Radio Observatory and the Caltech radio group, including Kristen Bernasconi, Stephanie Cha-Ramos, Sarah Harnach, Tom Klinefelter, Lori McGraw, Corey Posner, Andres Rizo, Michael Virgin, Scott White, and Thomas Zentmyer. Their tireless efforts were instrumental to the success of the DSA-110. The DSA-110 is supported by the National Science Foundation Mid-Scale Innovations Program in Astronomical Sciences (MSIP) under grant AST-1836018. We acknowledge use of the VLA calibrator manual and the radio fundamental catalog. Some of the data presented herein were obtained at the W. M. Keck Observatory, which is operated as a scientific partnership among the California Institute of Technology, the University of California and the National Aeronautics and Space Administration. The Observatory was made possible by the generous financial support of the W. M. Keck Foundation.

This research has made use of the NASA/IPAC Extragalactic Database (NED), which is operated by the Jet Propulsion Laboratory, California Institute of Technology, under contract with the National Aeronautics and Space Administration. This research has made use of NASA’s Astrophysics Data System Bibliographic Services. This research has made use of the VizieR catalogue access tool, CDS, Strasbourg, France (DOI:10.26093/cds/vizier). The original description of the VizieR service was published in 2000, A\&AS 143, 23. This research made use of Astropy, a community-developed core Python package for Astronomy.

\end{acknowledgments}

\begin{figure*}
    \includegraphics[width=\textwidth]{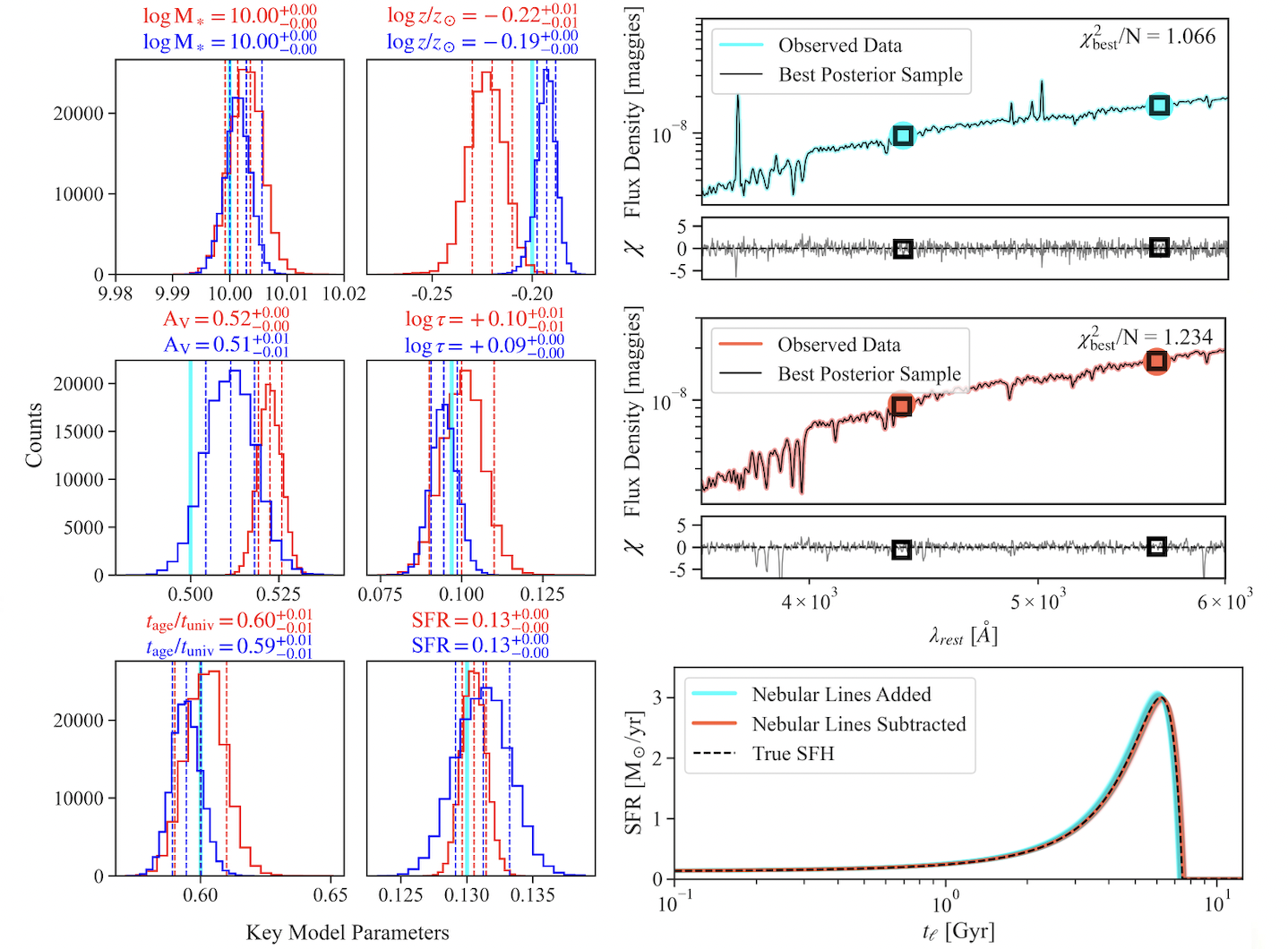}
    \caption{Comparison of different treatments for nebular emission in a galaxy. The left panel shows all the simulated true values (cyan), recovered values with nebular emission lines (blue), and recovered values with nebular emission lines removed (red). We observe that the constraints are better with emission lines, but both techniques yield broadly consistent values with the true values. The top (middle) right panel shows the simulated spectrum with (without) emission lines and recovered best posterior sample, along with the corresponding residuals. The bottom right panel shows the true and recovered star formation histories for both cases, which are consistent.}
    \label{fig:treating_neb_em}
\end{figure*}

\vspace{5mm}
\facilities{DSA-110, Keck:I (LRIS)~\citep{1995PASP..107..375O}}

\begin{figure*}
    \includegraphics[width=\textwidth]{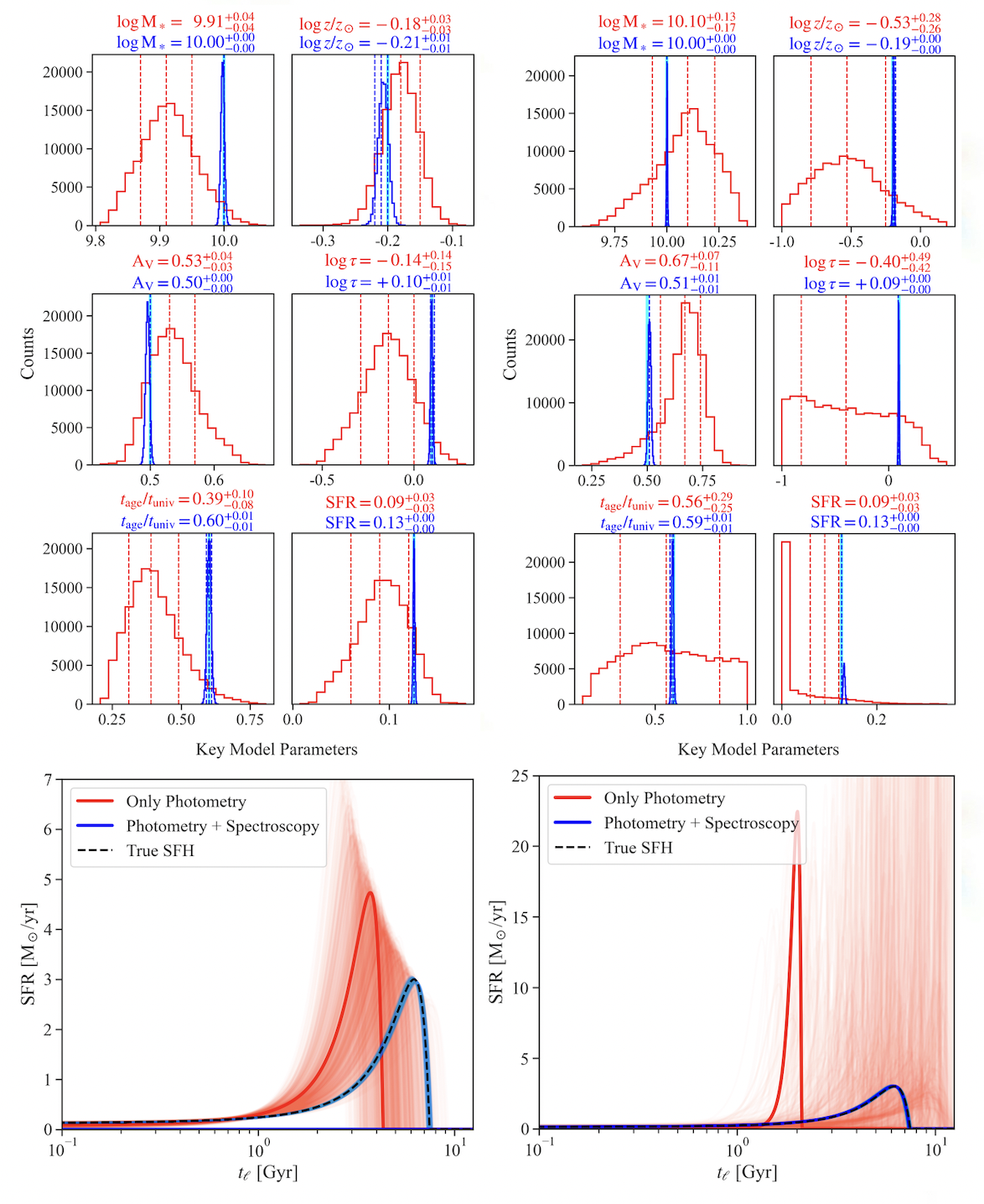}
    \caption{Demonstration of the constraints on star formation histories when fitting SED to photometry alone (red) and jointly fitting photometry and spectroscopy (blue). The left panel shows the results when nebular emission is omitted in the model for simplification. We observe that the star formation history constrained without spectrum is poor. On the other hand, an addition of nebular emission to the model complexify the problem, leading to even poor constraints without spectrum, as can be seen in the right panel. Hence, if accurately constraining star-formation history is important for a specific science case, we strongly recommend jointly fitting for photometry and high SNR spectrum in SED analysis.}
    \label{fig:parametric-spec-vs-no-spec}
\end{figure*}

\software{astropy \citep{2013A&A...558A..33A,2018AJ....156..123A}, {NumPy}~\citep{2020Natur.585..357H}, {SciPy}~\citep{2020NatMe..17..261V}, {Matplotlib}~\citep{2007CSE.....9...90H}, lpipe~\citep{2019PASP..131h4503P}, pPXF~\citep{2017MNRAS.466..798C, 2022arXiv220814974C}, Prospector~\citep{2021ApJS..254...22J}, emcee~\citep{2013PASP..125..306F}, astropath~\citep{2021ApJ...911...95A}, astro-datalab\footnote{\url{https://github.com/astro-datalab/datalab/}}}

\appendix

\section{SED Analysis Methodology}

\subsection{Treating Nebular Emission} \label{subsec:nebular_emission}

Modeling nebular emission in spectroscopic data can be tricky when fitting for photometry and spectroscopy together in \sw{Prospector}. In this appendix, we outline two approaches for tackling this. For the purpose of demonstrations, we simulate an SNR = 100 SED with parametric star formation history. We chose to use an exponentially declining star formation history parameterized by star formation timescale, $\tau = 1.25$ and age of the galaxy, $t_{\mathrm{age}} = 0.6 \times t_{\mathrm{univ}}$, where $t_{\mathrm{univ}}$ is the age of the galaxy at a redshift of $z_{\mathrm{red}} = 0.1$. The simulated photometry and spectrum represent a galaxy with a stellar mass M$_\ast = 10^{10}~\mathrm{M}_\odot$, metallicity $\log{z/z_\odot} = -0.2$, dust attenuation $A_V = 0.5$ and nebular ionization parameter $U_\mathrm{neb} = -3$. The true value of recent SFR averaged over the last 100~Myr is $\sim$0.13~M$_\odot$/yr. 

To deal with nebular emission, we demonstrate two possible approaches. First, adding nebular continuum and emission lines to the model along with marginalization over the amplitude of the emission lines to account for nebular emission from mechanisms other than star formation and nebular emission from old stellar population, as is also discussed in \citet{2021ApJS..254...22J}. The second approach involves subtracting the emission lines from the spectrum using the best fit for the gas component from \sw{pPXF} and then fitting these emission lines subtracted spectrum and photometry in \sw{Prospector} with nebular continuum added to the model. 

The resulting recovered star formation histories from our two experiments are shown in Figure~\ref{fig:treating_neb_em}. The reduced-$\chi^2$ of the best posterior sample when including the nebular emission lines is relatively lower than the best posterior sample when removing the nebular emission lines. We observe higher $\chi$ values at the higher energy hydrogen absorption features since they are not included in the \sw{pPXF} fit to the gas component. The recovered galaxy parameters are broadly consistent with the true parameters with slight deviations in metallicity and dust attenuation from their respective true values. The parameters are better constrained when nebular emission lines are included in the data. Nevertheless, both techniques are equally good at recovering the true star formation history and one may opt for either of these methods when modeling their respective galaxies.

\subsection{Constraining Star Formation History} \label{subsec:sfh_constraints}

The star formation history carries the information of the times of stellar birth in a galaxy and constraining it is important to achieve meaningful constraints on the delay-time distribution. The nebular emission and absorption features in a spectrum carry detailed information about the stellar age. In this appendix, we demonstrate how accurately one can recover the star formation history with and without using spectrum in SED fits, in the presence or absence of nebular emission. For the purpose of these demonstrations, we use the same galaxy parameters as used in Appendix~\ref{subsec:nebular_emission} and the same model as described in \S~\ref{subsec:sed_analysis} (except for changing the nebular emission based on the case under consideration).

The left panel of Figure~\ref{fig:parametric-spec-vs-no-spec} displays our results when nebular emission is completely excluded from our simulations. We observe that the recovered stellar mass, metallicity and dust attenuation are broadly consistent with the true values both, when we fit for photometry alone and fit simultaneously for photometry and spectroscopy. However, in the absence of spectrum, the age of the galaxy and the star formation timescale are poorly constrained, as can also be seen in the star formation history samples plotted in the bottom-left panel of Figure~\ref{fig:parametric-spec-vs-no-spec}. As was also noted in \citet{2021ApJS..254...22J}, we also observed the dust-age-metallicity degeneracy and stellar age-stellar age timescale degeneracy in our recovered parameters in the absence of the spectrum.

We further test this result by adding nebular emission to the model. We observe that the constraints of all parameters are poor when compared to the case with spectrum added to the SED fits. The stellar age and stellar evolution timescale parameters essentially recover the prior. This is also evident in the corresponding recovered star formation histories with and without spectrum in the bottom panel of Figure~\ref{fig:parametric-spec-vs-no-spec}. Based on these demonstrations, we strongly recommend using spectrum for constraining the star formation histories when possible, especially when doing delay-time distribution studies.



\bibliography{sample631}{}
\bibliographystyle{aasjournal}

\end{document}